# High rate tests of the photon detection system for the LHCb RICH Upgrade

M. P. Blago[a,b] and F. Keizer[c] on behalf of the LHCb RICH Collaboration

[a] *CERN, Geneva, Switzerland*
[b] *Ruprecht-Karls-Universität Heidelberg, Heidelberg, Germany*
[c] *University of Cambridge, Cambridge, United Kingdom*

## Abstract

The photon detection system for the LHCb RICH Upgrade consists of an array of multianode photomultiplier tubes (MaPMTs) read out by custom-built modular electronics. The behaviour of the whole chain was studied at CERN using a pulsed laser. Threshold scans were performed in order to study the MaPMT pulse-height spectra at high event rates and different photon intensities. The results show a reduction in photon detection efficiency at 900 V bias voltage, marked by a 20 % decrease in the single-photon peak height, when increasing the event rate from 100 kHz to 20 MHz. This reduction was not observed at 1000 V bias voltage.

*Keywords:* LHCb RICH Upgrade, 40 MHz event rate, MaPMT

## 1. Introduction

The LHCb detector upgrade aims to make more efficient use of the available luminosity during the CERN LHC Run III [1]. The photon detection system of the ring-imaging Cherenkov (RICH) detectors will be upgraded using two types of 8×8 channel MaPMTs: a 1×1 inch$^2$ ‘R-type’ to be used in high occupancy regions and a 2×2 inch$^2$ ‘H-type’ to be used in low occupancy regions. The MaPMTs are coupled to a baseboard, which contains the bias voltage bleeder chain and routes the signals to the CLARO ASICs [2] for pulse shaping and discrimination. The thresholds for discrimination are set on a channel-to-channel basis. Eight CLAROs are positioned on each front-end board (FEB). The digital boards configure the CLAROs and process events into Ethernet packets sent to the data acquisition (DAQ) system. Four R-type MaPMTs or one H-type MaPMT, as shown in Figure 1, constitute an elementary cell (EC) [3].

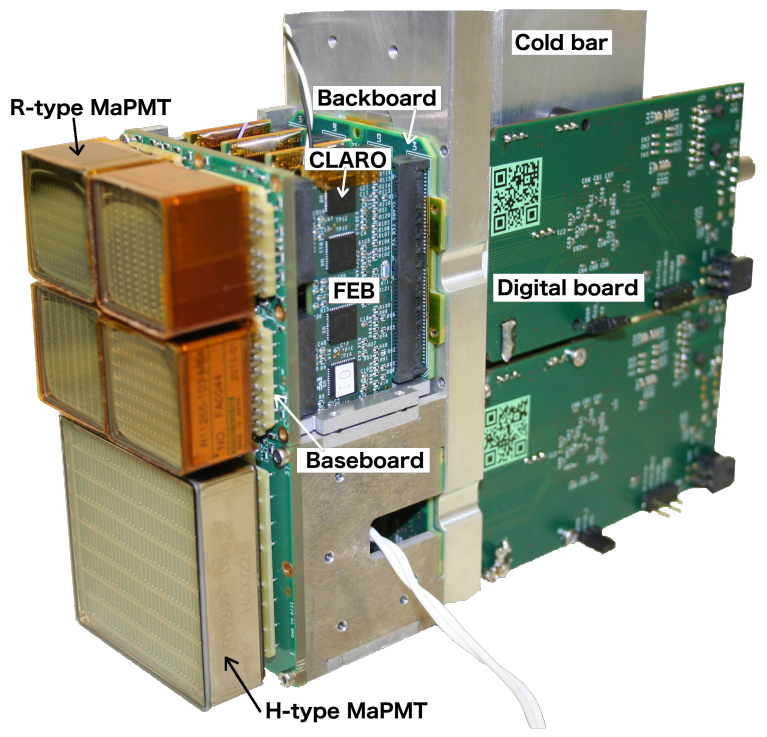

Figure 1: Picture of the coldbar with R-type (top) and H-type (bottom) elementary cells and digital boards mounted.

Key challenges for the photon detection system include single Cherenkov photon sensitivity, the 40 MHz event rate of the LHC proton-proton collisions and regions with high photon occupancy. A laser-based test facility was set up at CERN to investigate the performance of the MaPMTs and readout electronics under such conditions. The tests described in this paper complement those performed in a charged particle beam [4].

## 2. Experimental method

A schematic of the test setup is shown in Figure 2. The system was placed inside a light-tight box and studied using a pulsed laser with 405 nm wavelength. The pulse rate and the light intensity were controlled at the external laser driver. The laser was aligned with the centre of the EC, producing non-uniform illumination of the MaPMTs.

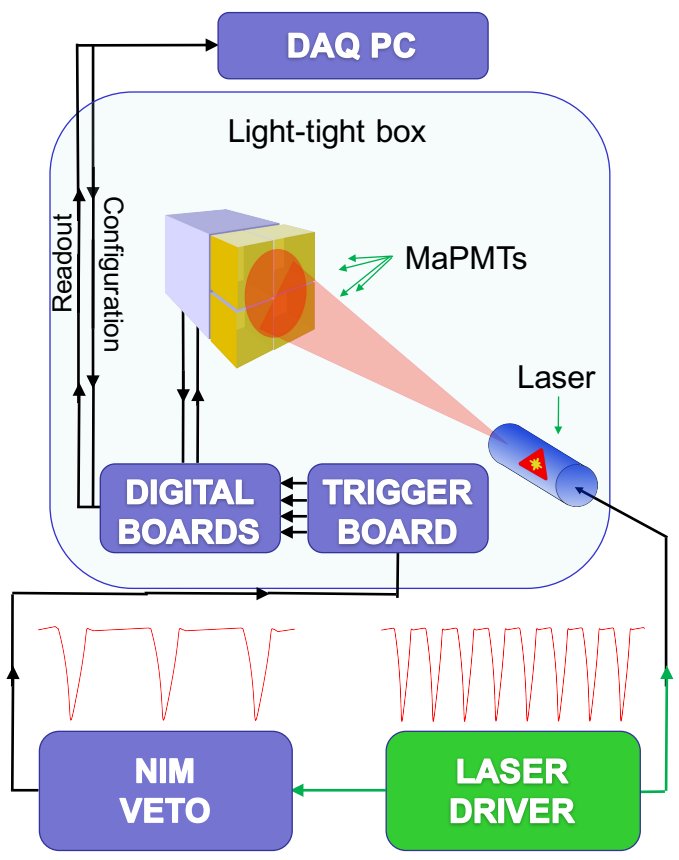

Figure 2: Schematic of the test setup. The MaPMTs are illuminated by a pulsed laser inside a light-tight box. A veto is used to reduce the data rate with respect to the event rate.

The digital boards were triggered from the laser driver with a veto applied to reduce the data rate to 20 kHz, which keeps the bandwidth within the limit of the Ethernet link between the digital board and DAQ PC. Recording only a subset of events does not affect the test results for the MaPMT and CLARO performance. The trigger board distributes the trigger to up to four digital boards. The MaPMTs were biased using a single high-voltage channel. The current drawn from this power supply channel was monitored in order to estimate the photoelectron current in the MaPMTs.

The pulse-height spectra of the MaPMTs were obtained using threshold scans. Every 250,000 readout triggers, the threshold is incremented by approximately 35,000 electrons (defined as 1 unit) at the CLARO input. The functionality of the CLARO to offset the thresholds by 32 units was used to study the pedestal characteristics in addition to the photon signals.

A typical threshold scan for three R-type MaPMT pixels is shown in Figure 3. At low threshold, the CLARO outputs are always high. The pedestal position is marked by a sharp drop in the fraction of events above threshold. At thresholds above the pedestal, the fraction of events is dominated by the integrated photon signal and was therefore used to estimate the photon occupancy of the channel, which is different for each pixel due to the non-uniform laser illumination.

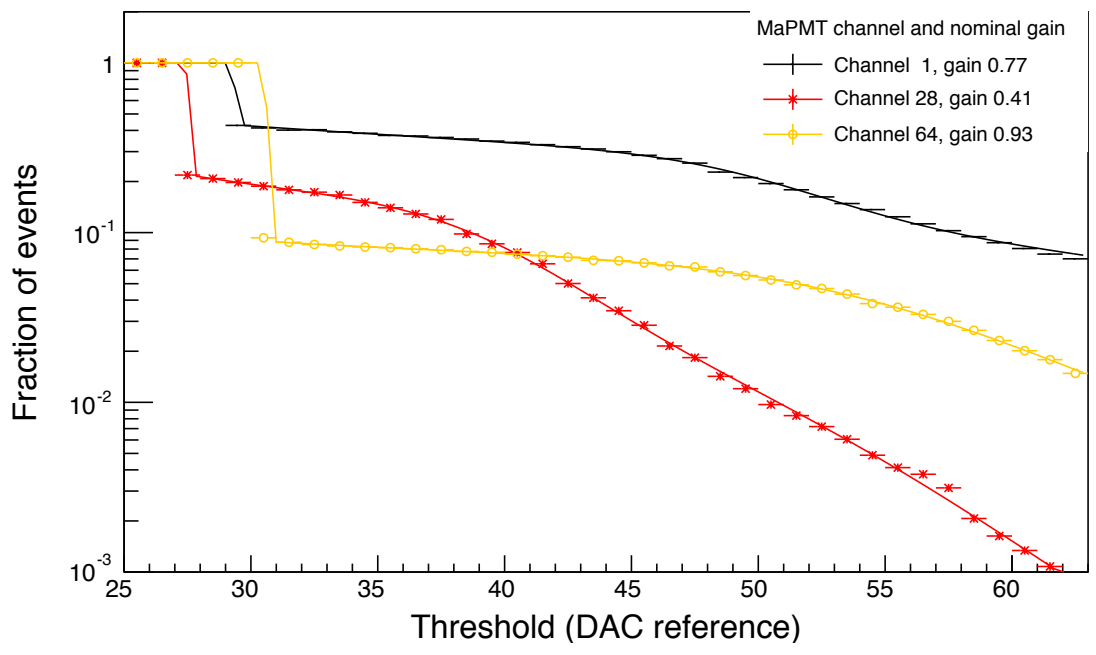

Figure 3: Threshold scans for three MaPMT channels with different gain and photon occupancy. The data are fitted in order to obtain the pulse-height spectra.

A function based on three Fermi probability density functions, representing the pedestal, first and second photon peak, as well as on an exponential to describe the right-hand side tail of the pedestal, is fitted to each threshold curve. The fit is then differentiated in order to obtain the MaPMT pulse-height spectrum, which contains information about the noise, gain and photon detection efficiency of the channel.

## 3. Results

### *3.1. R-Type MaPMT*

Figure 4 shows the percentage photon occupancy per channel of the R-type EC at 'low intensity'. The occupancy reaches nearly 30 % at the centre of the laser spot. For the high intensity tests, the central region has 45 % occupancy. The pulse-height spectra for R-type MaPMT channels at 900 V and 1000 V bias are shown in Figure 5a and 5b respectively. The effect of increasing the laser rate from 100 kHz to 20 MHz is shown at low and high photon occupancy. The observed effect is representative for all channels and is more pronounced at higher photon occupancy.

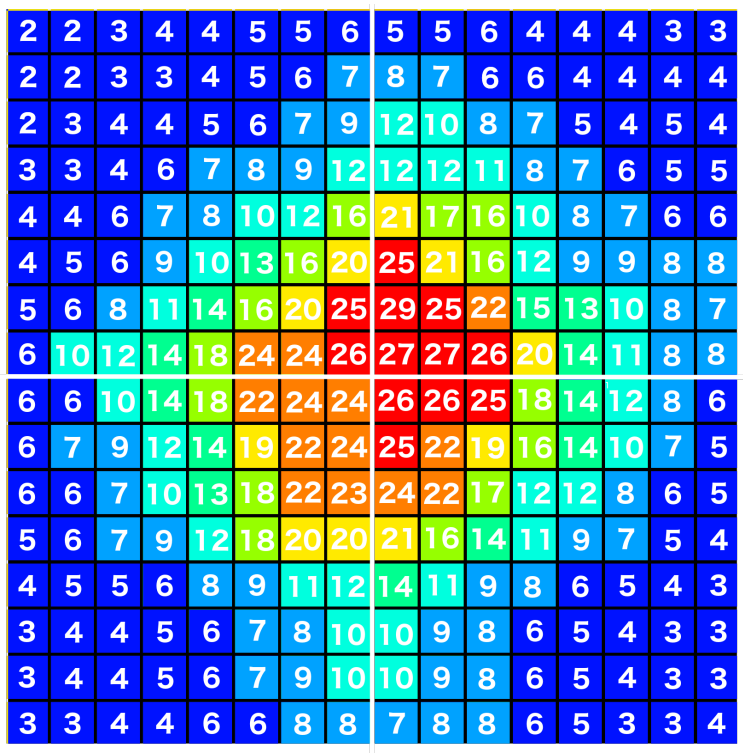

Figure 4: Photon occupancy (%) map for an elementary cell containing four R-type MaPMTs at low intensity setting. The occupancy decreases towards the edges of the elementary cell due to non-uniform illumination.

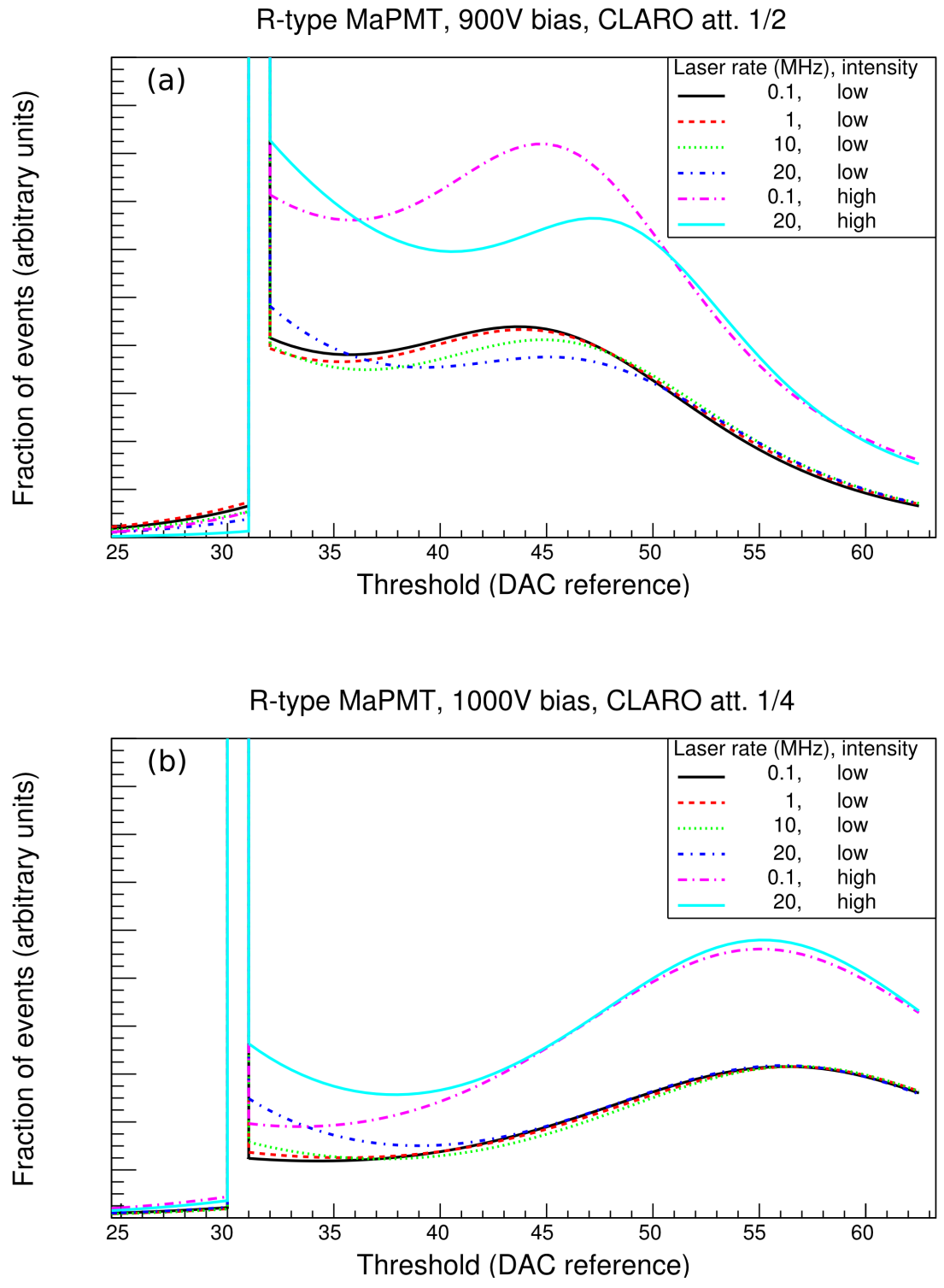

Figure 5: Pulse-height spectra of an R-type MaPMT pixel for different pulse rates and photon occupancies, at 900 V (a) and 1000 V (b) bias voltage.

The position of the pedestal does not vary as a function of the laser pulse rate or photon occupancy. At 900 V MaPMT bias, an increase of the pulse rate shifts the single photon peak position to a higher threshold value and reduces the peak height by approximately 20 %, indicating a lower photon detection efficiency. This can be caused by a reduction in the bias voltage at the last MaPMT dynode, leading to space-charge effects that lower the charge collection efficiency at the anode. The effect can be reduced by separately biasing the last two stages of the MaPMT dynode chain. At 1000 V, the space-charge effects become negligible and no reduction in photon detection efficiency is observed.

### *3.2. H-Type MaPMTs*

Figure 6 shows the percentage photon occupancy map of the H-type MaPMT. This photon occupancy is well above the expected value of less than 3 % in the H-type region of the RICH detector[1]. Figure 7 shows the MaPMT pulse-height spectra at 900 V, which contain a small reduction in the photon detection efficiency with increasing pulse rate. As for the R-type, the H-type MaPMTs show no significant change in photon detection efficiency or gain at high pulse rates at 1000 V bias. It can be concluded that the space-charge effects are less strong in the H-type than in the R-type MaPMT.

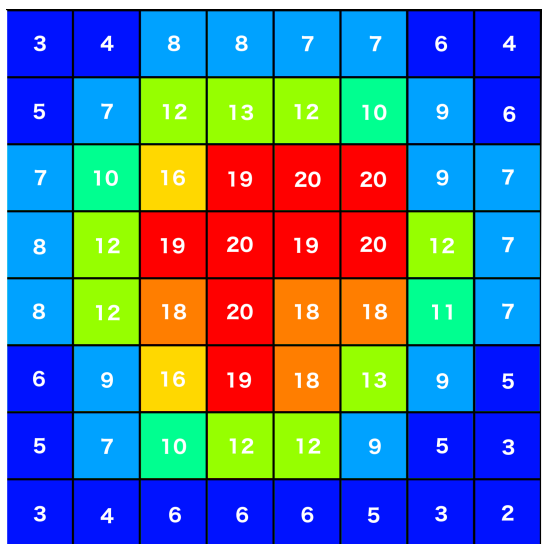

Figure 6: Photon occupancy (%) map for an H-type MaPMT.

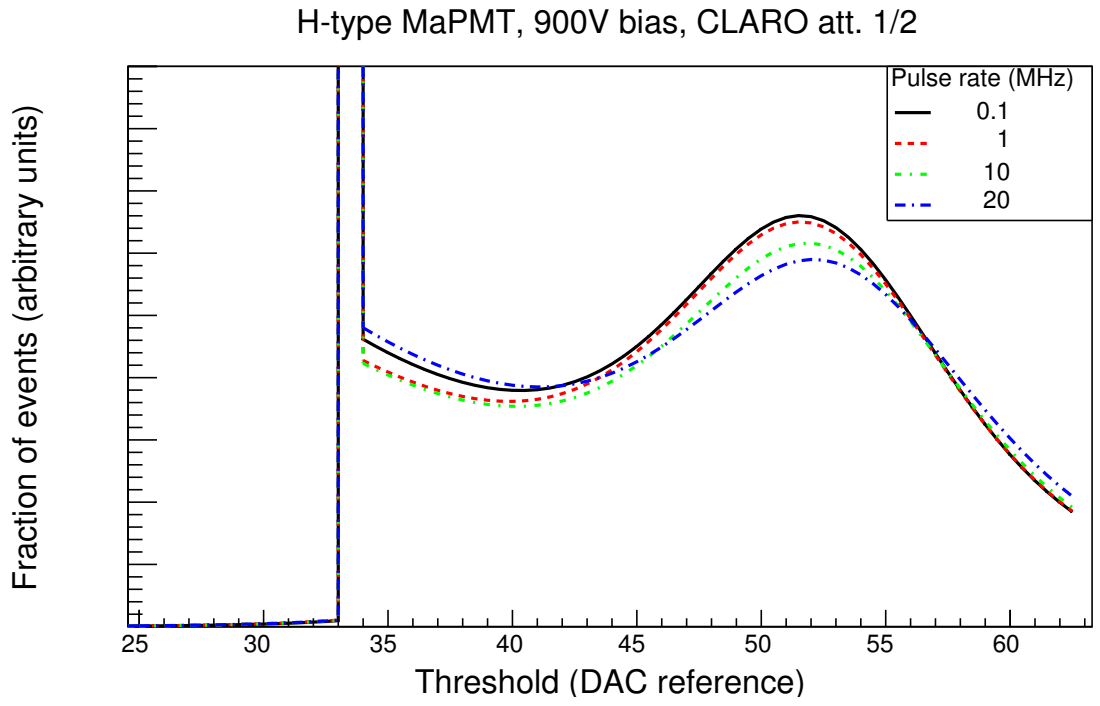

Figure 7: Pulse-height spectra of an H-type MaPMT pixel at 900 V bias voltage.

### *3.3. Photocathode Current*

The current drawn by the MaPMTs from the power supply was monitored as a function of the pulse rate, and converted into a photocathode current using a SPICE [5] simulation of the baseboard bleeder chain. Figure 8 shows the photocathode current as a function of the pulse rate. The measurements agree within error with a linear increase of the photocathode current from 100 kHz pulse rate, where the current is negligible, to 40 MHz, where it is close to 100 pA. In contrast to the threshold scans, where the sampling of the CLARO outputs by the digital board was limited to 20 MHz, the measurements of the photocathode current also include the 40 MHz rate.

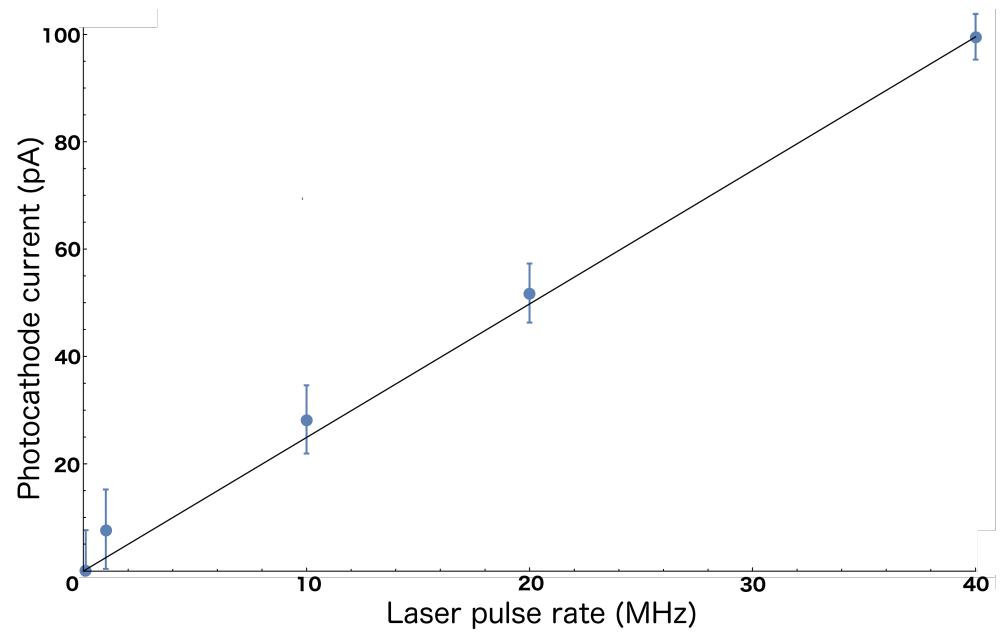

Figure 8: Plot of the photocathode current, calculated from the power supply current, as a function of laser rate up to 40 MHz. The data were taken with an R-type MaPMT at 1000 V bias voltage and at low laser intensity.

## 4. Conclusion and Further Tests

The LHCb RICH Upgrade photon detection system has been tested at pulse rates up to 20 MHz and photon occupancy up to 45 %. A reduction in photon detection efficiency was observed at high pulse rate at 900 V bias, which will be further investigated with the last MaPMT dynode powered separately. Future tests will involve the implementation of customised firmware for the digital boards, which allows the MaPMTs and readout electronics to be tested at 40 MHz rate.